\documentclass[fleqn,twoside,twocolumn,nofootinbib]{revtex4} 
\usepackage[sec]{ujp} 
\begin{document}
\title[OPTICAL QUALITY CHARACTERIZATION OF KDP CRYSTALS]
{OPTICAL QUALITY\\ CHARACTERIZATION OF KDP CRYSTALS\\ WITH
INCORPORATED TiO\boldmath$_2$ NANOPARTICLES\\
AND LASER SCATTERING EXPERIMENT SIMULATION}%
\author{V.Ya. GAYVORONSKY}
\author{V.N. STARKOV}
\author{M.A. KOPYLOVSKY}
\author{M.S.~BRODYN}
\author{E.A. VISHNYAKOV}
\author{A.Yu. BOYARCHUK}
\affiliation{Institute of Physics, Nat. Acad. of Sci. of Ukraine}
\address{46, Nauky Prosp., Kyiv 03680, Ukraine}
\email{vlad@iop.kiev.ua}
\author{I.M. PRITULA}%
\affiliation{STC ``Institute for Single Crystals'', Nat. Acad. of Sci. of Ukraine}%
\address{60, Lenin Ave., Kharkiv 61001, Ukraine}%
\udk{535.361.12+53.072}  \razd{\seciv}

\setcounter{page}{875}%
\maketitle

\begin{abstract}
We study the elastic scattering of light in pure KDP crystals and
KDP crystals with incorporated titanium dioxide nanoparticles. It is
shown that the optical quality of the crystals decreases
insufficiently for the used concentrations of nanoparticles. A
mathematical model of the experimental setup for light scattering
measurements in low-dispersion media is developed and discussed. The
propagation function of the experimental setup is given in
analytical form. The relevance of the model is verified with the use
of experimental scattering data.
\end{abstract}

\section{Introduction}

The paper is devoted to the optical quality characterization of the
novel nonlinear optical (NLO) material -- potassium dihydrogen
phosphate single crystal (KH$_2$PO$_4$, KDP) matrix with
incorporated titanium dioxide nanocrystals (anatase modification)
[1]. The hybrid material was designed for the phase matching
conditions and the second harmonic generation efficiency control by
light due to the resonant excitation of the surface states of
nanoparticles [2]. We applied the cone-shaped interference method to
study the effect of TiO$_2$ nanoparticles on the linear optical
response of the medium. A special attention was paid to investigate
the angular distribution of scattered light (scattering indicatrix)
and scattering extinction losses for a highly transparent
low-dispersion NLO functional material.

Experimental data on scattering in low-dispersion materials are
within a high dynamic range (6--7 orders of the signal magnitude).
To enable the measurements of weak signals of the scattered light,
we use a collecting lens on the registration CCD array with the
angular aperture that essentially exceeds the initial divergence of
a laser beam. This leads to an enhancement of the signal-to-noise
ratio at large scattering angles and simultaneously to the angular
averaging of registered data. The introduction of the experimental
setup apparatus function gives possibility to derive the precise
angular signal distribution using a proper mathematical processing
of experimental data.

Previously, a mathematical model was developed in [3]. It was shown
that the registered scattering signal is distorted at small
scattering angles. That model can be applied to high-dispersive
media like that described in [4], but it is insufficient for
low-dispersion samples due to the lacking of precision. For this
reason, a more precise model of the experimental setup was
developed. The consistency of the proposed model is verified, by
using experimental measurements in series of the single crystal KDP
matrix with different concentrations of TiO$_2$ nanocrystals [1].

\section{Experimental Setup}

For isotropic media, the scattering is axially symmetric, so it is
sufficient to measure a cross-section of the light power/intensity
angular distribution in any scattering plane (Fig.~1).

\begin{figure}
\includegraphics[width=8cm]{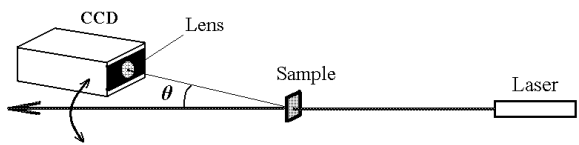}
\vskip-4mm\caption{Scheme of the experimental setup for angular
dependence measurements of the scattered light. The lens mounted at
a CCD array controls the solid angle for the scattering radiation
collection, $\theta$ denotes the angle between the incidence laser
beam and scattering signal acquisition directions }
\end{figure}

The  experimental laser setup for the optical diagnostics of
low-dispersion samples is based on a G-5 goniometer with different
laser sources on its fixed arm: a DPSS diode laser (wavelength
$\lambda$~= 532~nm, radiation power $P$~= 50~mW) or a He-Ne laser
($\lambda$~= 633~nm, $P$ = 10~mW). The sample under study is placed
at a sample stage, and the registration unit is mounted on the
movable arm of a goniometer at the distance $L$ = 200~mm from the
sample. It consists of a CCD array AMKO LTI MuLTIray (1024 pixels
with $25\times 200~\mu$m$^2$ pixel size, 12-bit digital resolution)
with a lens (diameter $d$~= 9.6~mm) attached to it. The rest
receiving area of the CCD is completely blackened, so that only the
light passed through the lens reaches the sensitive surface of the
CCD array.

The experimental setup is intended for the investigation of a polar
angular dependence of the scattered light power against to the
incidence laser beam propagation direction. The magnitudes of
radiation power along the registration direction $\theta$ in the
given solid angle were obtained via the summation of the all pixels
data of the CCD array.  Due to the focusing lens, the signal on CCD
is registered on a small number of pixels ($\sim 30$) in each
measurement, which extremely increases the signal-to-noise ratio.

The indicatrix registered with the use of this experimental setup is
different from the real one due to the spatial averaging with the
lens aperture. So it is reasonable to develop a mathematical route
to restore the original angular distribution from the registered
one. The model presented here is verified in a direct way, and it
can be a base for solving the inverse problem in future.

Scattering data with high angular resolution are required for the
verification of the model. This was achieved due to a reduction of
the solid angle for the radiation collection. The CCD array with a
wide aperture lens was replaced with a fiber coupled CCD
spectrometer with the input fiber aperture ($d = 410~\mu $m).
Unfortunately, the angular resolution enhancement leads to a drastic
reduction of the scattering signal level and the angular range of
data acquisition ($ \theta \thicksim \pm 2^\circ$).

The spectroscopic and scattering measurements were done with
crystalline plates $10 \times 10 \times 0.8$~mm$^3$ cut
perpendicular to the optical axis ($Z$-cut) of the uniaxial KDP
matrix. A set for the investigation consists of a pure KDP single
crystalline plate and KDP crystals with incorporated TiO$_2$
nanoparticles at different concentrations ($10^{-5}$, $10^{-4}$, and
$10^{-3}$ wt.\% TiO$_2$) in a mother liquor solution. There were two
series KDP(Pr) and KDP(P) cut from the prismatic $\{100\}$ and
pyramidal $\{101\}$ growth sectors, respectively. The pure KDP
crystals will be denominated simply as P or Pr, and the rest of
samples will be further designated according to the nomenclature in
Table~1, where we also give the thickness $h$ of each sample. In
some cases, the TiO$_2$ nanoparticles concentration is written
explicitly in a short form (e.g., $10^{-3}$ wt.\%), which is
referred to the nanoparticles concentration in the mother liquor
solution.

Due to the opposite charge signs of the termination interfaces of
the different growth sectors, the $\{101\}$ (P) growth sector
captures more TiO$_2$ nanocrystals in comparison with the $\{100\}$
(Pr) sector at the growth stage. The preliminary research has shown
that the average concentration of incorporated nanocrystals into the
KDP matrix is twice less in comparison with the nanoparticles
concentration in the mother liquor solution.

\begin{table}[b]
\noindent\caption{Spectral properties of the samples (measured with
a fiber coupled spectrophotometer). ``Pr''  denotes the prismatic
\boldmath$\{100\}$ growth sector, ``P'' --- pyramidal $\{101\}$.
TiO$_2$ concentrations are given for mother liquor solutions. $h$ is
the crystalline plate thickness}\vskip3mm\tabcolsep=6.7pt

\noindent{\footnotesize\begin{tabular}{ccccc}
  \hline
        \multicolumn{1}{c}{\rule{0pt}{9pt}TiO$_2$ conc., }& \multicolumn{1}{|c}{Sample} & \multicolumn{1}{|c}{$h$,} & \multicolumn{2}{|c}{Transmittance, \% }
        \\ \cline{4-5}
        \multicolumn{1}{c}{wt.\% }& \multicolumn{1}{|c}{notation} & \multicolumn{1}{|c}{mm} & \multicolumn{1}{|c}{$\lambda=532$ nm} & \multicolumn{1}{|c}{$\lambda=633$ nm} \\ \hline

        $0$ & P & 0.81 & $91.3 \pm 1.0$ & $91.9 \pm 0.9$ \\

        & Pr & 0.83 & $92.4 \pm 0.6$ & $93.0 \pm 0.7$ \\ [1.5mm]

        $10^{-5}$ & P-5 & 0.75 & $89.8 \pm 0.9$ & $90.2 \pm 1.1$ \\ 

        & Pr-5 & 0.78 & $90.3 \pm 0.9$ & $91.0 \pm 1.0$ \\ [1.5mm] 

        $10^{-4}$ & P-4 & 0.76 & $89.8 \pm 1.2$ & $90.8 \pm 1.0$ \\ 

        & Pr-4 & 0.75 & $91.4 \pm 1.1$ & $92.0 \pm 1.0$ \\[1.5mm] 

        $10^{-3}$ & Pr-3 & 0.81 & $89.1 \pm 0.5$ & $89.7 \pm 0.5$ \\ \hline
 \end{tabular}}
\end{table}

\section{Mathematical Model, Experimental Results and Discussion}

\subsection {Optical quality characterization of crystals}

\subsubsection*{3.1.1. Spectral transmittances}

Optical transmittance spectra of the samples were measured with a
Perkin Elmer Lambda 35 spectrometer in the range 200--1100~nm (the
reference is air). The spectra were studied for the samples with
polished (001) faces $\sim $0.8~mm in thickness without
anireflective coatings.

The results of these spectral measurements are presented in Fig.~2,
and the samples notations are given in Table~1. The transmittance
magnitudes in the visible and near-IR ranges are governed, in
principle, by the reflections from crystalline-air interfaces.

It is known that that the pure KDP crystals have different
transmittances for prismatic (Pr) and pyramidal (P) growth sectors
in the UV spectral range [1]. The P samples are more transparent due
to a less efficient absorption of impurities from the mother liquor.
In Fig.~2, we show a typical transmittance reduction $\sim 75~\%$ at
$\lambda $~= 270~nm (curve {\it 1}). Pr-5 and Pr-3 (curves {\it 2}
and {\it 4}) have transmittances similar to that of the pure one in
the range 260--350~nm, while Pr-4 (curve {\it 3}) demonstrates a
much higher transmittance in comparison with those of the rest ones.
It gains $\sim 88$\% at 300~nm. The magnitude is comparable with the
transmittance of the pure P crystal, and it exceeds the transmission
of the pure Pr one by about 8\%.

We suggest that the mentioned peculiarity can be explained due to
the compensation effect of incorporated nanoparticles. For the
prismatic growth sector, a concentration of $10^{-4}$ wt.\% TiO$_2$
is not so high for the efficient absorption of UV light by anatase
nanoparticles due to the direct transitions from the valence band to
the conduction one. However, the concentration of incorporated
nanoparticles is enough to absorb impurity atoms, especially the
atoms of polyvalent metals on their developed surface during the
growth
stage.

We have also tested a homogeneity of the optical transmittances of
crystals over the transverse plane with an Ocean Optics type fiber
coupled spectrophotometer in the range 450--900~nm. The samples were
illuminated with a broadband collimated light beam, and the readout
of transmittances was performed with a microobjective attached to
the fiber. It was placed at a distance of 2~mm from the crystal
output interface, so it was possible to collect almost all the
transmitted photons (both ballistic and diffusive ones). We
performed 6--10 measurements for each sample. Their averaged
transmission coefficients at 532 and 633~nm are presented in
Table~1.

\begin{figure}
\includegraphics[width=\column]{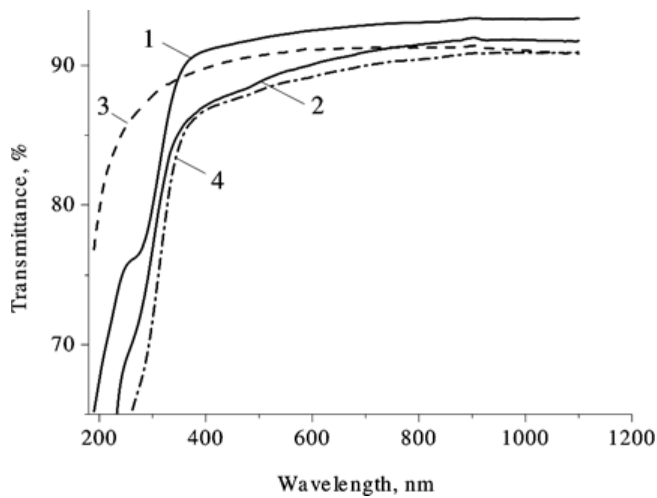}
\caption{Spectral transmittances of KDP single crystals with
incorporated TiO$_2$ nanoparticles cut from the prismatic growth
sector (Pr):  {\it 1} -- pure KDP (Pr), {\it 2} -- KDP:TiO$_2$ with
$10^{-5}$ wt.\% (Pr-5), {\it 3} -- $10^{-4}$ wt.\% (Pr-4), {\it 4}
-- $10^{-3}$ wt.\% (Pr-3) samples}
\end{figure}

We selected the wavelengths for the characterization of optical
homogeneity and scattering properties by the following reasons. The
available laser sources at the mentioned wavelengths correspond to
different regimes of the excitation of pure and hybrid materials.
The wavelength $\lambda $~= 633~nm corresponds to the nonresonant
excitation of the KDP matrix and TiO$_2$ nanoparticles. The
wavelength $\lambda $~= 532~nm leads to the resonance-type
excitation for intrinsic matrix defects [5] and to the electron
excitation from the valence band into the deep level of a defect in
the gap (oxygen vacancies) of anatase nanoparticles~[2].\looseness=1

The analysis of the data presented in Table~1 has shown that the
nonresonant transmittance is a little bit higher than that in the
resonance excitation case. A dispersion of the optical inhomogeneity
is the most evident for the resonant excitation regime ($\lambda$~=
532~nm) and for the $10^{-4}$ wt.\% TiO$_2$ concentration. This can
be explained by a high sensitivity of the compensation effect to the
local concentration of anatase nanoparticles in the matrix.

The comparison of the data in Table~1 with the transmittance curves
in Fig.~1 obtained with two different spectroscopic setups shows the
high optical transmittance and the homogeneity of pure KDP and novel
hybrid KDP:TiO$_2$ single crystals in the visible and near-IR
ranges. We observed the $\sim 1\%$ difference of the transmission
coefficient for the investigated samples in the visible range. In
the conventional mode of a Perkin Elmer spectrometer (without
integrated sphere detection), the samples were placed at a distance
of 200~mm from the $1 \times 10$~mm$^{2}$ slit of a detector unit.
We explain the obtained minor magnitudes of the Perkin Elmer
spectroscopic data by the presence of scattering losses, which was
proved in the experiment dealing with scattering properties of the
novel hybrid material.

\begin{figure*}
\includegraphics[width=17cm]{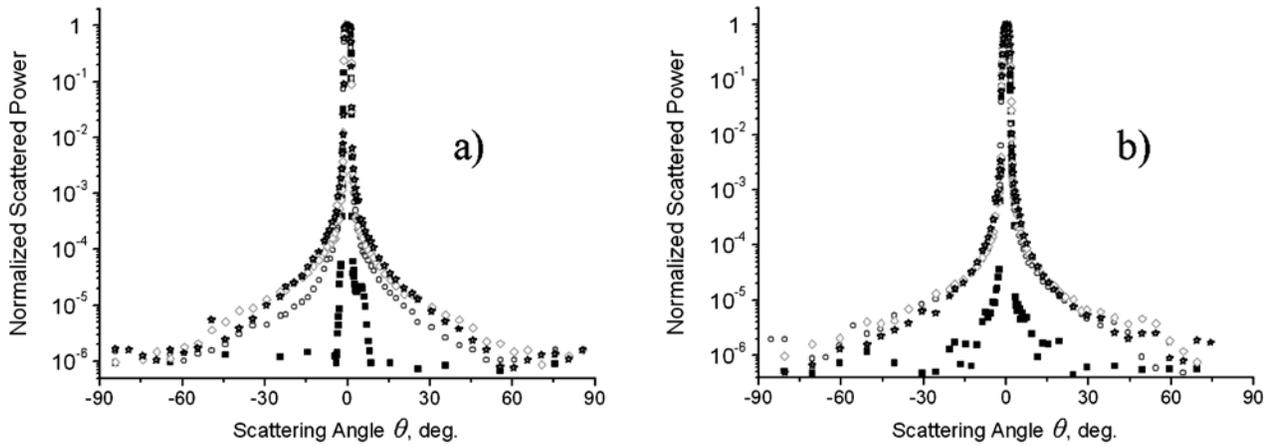}
\vskip-3mm\caption{Scattering indicatrices on the semilogarithmic
scale for KDP(P) crystals series at ({\it a}) 532-nm and ({\it b})
633-nm wavelengths. Squares  -- freely propagating laser beam;
$\circ$ -- pure P crystal; $\diamond$ -- KDP:TiO$_2$ P-5 sample;
$\star$ -- KDP:TiO$_2$ P-4 sample}
\end{figure*}

\begin{figure}
\includegraphics[width=\column]{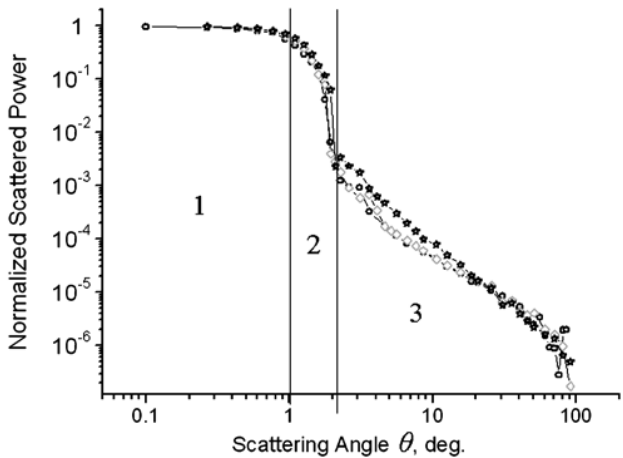}
\vskip-3mm\caption{Scattering indicatrices for the series of KDP(P)
single crystals on the double logarithmic scale at 633~nm.  $\circ$
-- pure P crystal, $\diamond$ -- KDP:TiO$_2$ P-5 sample, $\star$ --
P-4 sample}
\end{figure}

\subsubsection*{3.1.2. Optical scattering properties}

We have studied the angular distribution of scattered light for both
series of KDP(Pr) and KDP(P) crystals in two regimes of excitation:
(a) resonant regime at 532~ nm and (b) nonresonant one at 633~nm.
The measurements were carried out only in the forward hemisphere,
i.e. in the angle range $0 \leqslant \theta \leqslant 90^{\circ}$.
The laser beams propagated along the optical axis ($\theta = 0$) for
$Z$-cut single crystals. We performed the axial rotation of the
crystalline plates and observed the same angular distribution of the
scattered light (isotropic character of the effect).

The measured scattering indicatrices for the series of KDP(P)
crystals are presented at Fig.~3. Squares corresponds to the angular
distribution of the scattered radiation of the laser beams
propagating in air at 532~nm ({\it a}) and 633~nm ({\it b}). The
data are presented on the semilogarithmic scale in order to cover a
huge dynamic range of the described process -- about 6--7 orders of
the signal magnitude reduction at large scattering angles. All the
data are normalized on the total transmitted power $P_0$ in the
forward hemisphere. It was estimated by the incident power of the
laser sources and the total transmittance data presented at Table~1.

Due to the spatial averaging effect of the lens attached to the CCD
array, all the curves look like very similar: a sharp bell-shaped
part and wings. In order to understand the physical origin of each
part, we present the data for the KDP(P) series on the double
logarithmic scale (see Fig.~4). One can clearly see how the presence
of the lens distorts the shape of the registered indicatrix. One can
split each curve into three typical ranges, where the investigated
dependences are distorted at different extents.

The first one ($\theta \leqslant 1^{\circ}$) is a plateau. In this
``blurred'' interval, the lens collected almost the whole laser beam
transmitted through the sample -- the so-called ballistic photon
registration geometry. Thus, the acquired data do not depend on the
angular position of the lens. It is the range of the biggest
distortions, because, according to the spatial averaging, no
peculiarities in the indicatrix can be resolved at such small
angles.

In the second interval $1^{\circ} \leqslant \theta \leqslant
2^{\circ}$, an abrupt decrease of the registered signal is observed,
when the lens departs from the transmitted laser spot area. This
corresponds to the direct light to shadow angular range scan with
large experimental errors (up to $20\%$) due to the angular position
uncertainty $\Delta \theta$ of the measuring unit.

In the third interval $\theta \geqslant 2^{\circ}$, the lens
collects only the light scattered by a sample. For a medium without
spatially periodic structures, the angular distribution of the
scattered light is smooth on the logarithmic scale. Thus, the
spatial averaging of the collecting lens has no essential impact on
the angular resolution of a registered signal. A slight reduction of
the angular resolution is compensated by the essential enhancement
of a registered signal level.

In the third interval (Fig.~4), we observed an almost linear
decrease of the registered signal on the double logarithmic scale.
This corresponds to the wing parts of the same dependences plotted
on the semilogarithmic scale in Fig.~3. The slope $-2.0 \pm 0.3$
($2^{\circ} \leqslant \theta \leqslant 45^{\circ}$) is the same for
the different series of samples and for the different readout
wavelengths. This means that the scattered radiation power along the
$\theta$ direction obeys the inverse quadric law $\Delta P (\theta)
\sim \theta^{-2}$.

Due to negligible distortions of the angular distribution of
scattering data in the mentioned intervale, the registered
indicatrix is very similar to the real one. This enables us to
estimate the scattering losses in the forward hemisphere $p_{\rm
scat}$ for all samples under study. The method of calculation is
based on the integration of the scattered light power $\Delta
P(\theta)$ in the solid angle $\Delta \Omega$ over the part of the
forward hemisphere that corresponds to the 3$^{{\rm rd}}$ scattering
area in Fig.~4,
\begin{equation}
 p_{\rm scat} = \frac{2 \pi}{P_0} \int \limits_{\theta_{\rm min}}^{\pi/2}
 \frac{\Delta P(\theta)}{\Delta \Omega} \sin \theta d\theta,
\end{equation}

\noindent where the lens incoming solid angle $\Delta \Omega = 4 \pi
\sin^2{\theta_{\rm lens}/2} = 1.8 \times 10^{-3}$~sr, $\theta_{\min}
\sim 1.8^{\circ}$ exceeds the lens angular aperture $\theta_{\rm
lens}=\arctg( d/2L ) \sim 1.4^{\circ} $ due to a finite size of the
transmitted beam at the registration plane. The data were normalized
on the total power $P_0$ transmitted into the forward hemisphere.

\begin{table}[b]
\noindent\caption{Scattering losses into the forward hemisphere for
KDP:TiO\boldmath$_2$  crystals series normalized on total power
$P_0$ transmitted into the forward hemisphere,
\%}\vskip3mm\tabcolsep12.2pt

\noindent{\footnotesize\begin{tabular}{ccccc}
  \hline
              & \multicolumn{1}{|c}{Pure}  & \multicolumn{3}{|c}{TiO$_2$ conc., wt.\%} \\
        \cline{3-5}

        & \multicolumn{1}{|c}{KDP}  & \multicolumn{1}{|c}{$10^{-5}$} & \multicolumn{1}{|c}{$10^{-4}$} & \multicolumn{1}{|c}{$10^{-3}$} \\
        \hline

        P (633 nm) & 1.0   & 1.7   & 1.8   & -- \\ 

        Pr (633 nm) & 1.1   & 2.6   & 2.6   & 2.6 \\ 

        P (532 nm) & 1.0   & 1.7   & 2.5   & -- \\ 

        Pr (532 nm) & 1.1   & 1.7   & 1.3   & 2.8 \\ \hline
 \end{tabular}}
\end{table}

The results of estimations of the scattering losses $p_{\rm scat}$
via Eq.~(1) are presented in Table~2. The pure P and Pr single
crystals scatter about $1\%$ of the light radiation for both laser
wavelengths. The ten-percent increase of the losses in the Pr
crystal in comparison with the P one can be attributed to a higher
concentration of impurities in the former [1]. The incorporation of
TiO$_2$ nanoparticles insignificantly reduces the optical quality of
the crystals. Even for $10^{-3}$ TiO$_2$ wt.\% concentration, the
scattering losses do not exceed $3\%$ that makes it possible to
utilize the hybrid crystals for the different optical applications.

In order to explain the obtained data on scattering losses, we
should refer to the structural characterization of the hybrid media.
The high-resolution X-ray diffraction analysis has shown that the
incorporation of nanoparticles has no essential effect on the
structural perfection of composite materials [1]. In the growth
process, the KDP matrix can capture nanoparticles between the growth
layer stacks, by effectively forming a 1D layered macroscopic
structure …KDP:TiO$_2$:KDP… with a spatial period of 20--30~$\mu$m.
This gives possibility to incorporate TiO$_2$ nanocrystals, whose
nonlinear optical response can be controlled by the resonant
excitation of surface states [2].

Thus, the anatase nanocrystals are distributed on the planes of the
boundaries of KDP growth layer stacks with a definite spatial
period. The orientations of these planes relative to the optical
axis are different for the P and Pr series. In the $Z$-cut plates,
the optical axis is parallel to the planes of boundaries with
TiO$_2$ nanoparticles for the Pr crystals, and it makes an angle of
about $44^\circ$ for the P one.

For the nonresonant readout case, the scattering losses at 633~nm in
both series of KDP(P) and KDP(Pr) crystals are almost independent of
the TiO$_2$ concentration. A reduction of the scattering losses by $
\sim 30$\% in the P series can be explained by the tilt of the
planes with TiO$_2$ nanoparticles toward the laser beam propagation
(optical axis) direction, which leads to a decrease of the
interaction length of light with nanoparticles.

In the case of the resonant excitation at 532~nm, the effect from
TiO$_2$ nanoparticles is evident. It is clearly seen in Fig.~3 for
the P series at the part with indicatrix wings. The scattering
losses at 532~nm ({\it a}) are higher for the crystals with
incorporated nanoparticles as compared with those for the pure one.
But, in the case of the 633-nm irradiation ({\it b}), they are much
alike. The noise level reference of the acquisition unit corresponds
to the wings of the laser beam scattering indicatrix.

The monotone growth of the scattering losses with the concentration
of nanoparticles for the P series is determined by the following.
The preliminary research has shown that the KDP:TiO$_2$ single
crystals manifest a self-focusing effect with the CW laser
excitation at 532~nm due to the resonant excitation of anatase
nanoparticles. The effect leads to the optical contrast enhancement
in a vicinity of the nanocrystal and to an increase of its
scattering cross-section.

\begin{figure}
\includegraphics[width=\column]{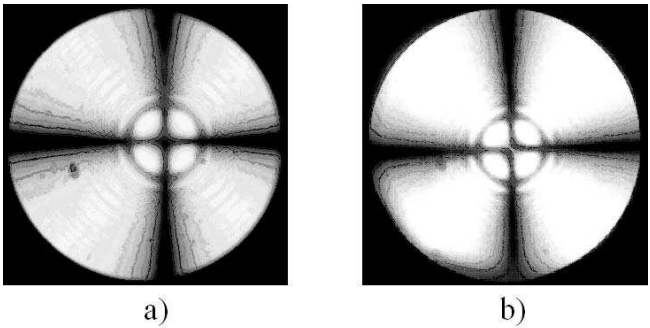}
\vskip-3mm\caption{Cone-shaped interference images for $Z$-cut
plates of ({\it a}) pure KDP(P), ({\it b}) KDP:TiO$_2$ (P-4) single
crystal plates }
\end{figure}

The same effect in conjunction with a specific alignment of
nanoparticles toward the beam propagation direction produces a more
complicate response to a variation of the concentration of
nanoparticles in the Pr series. An increment of the refractive index
at the boundaries of KDP growth layer stacks with TiO$_2$
nanoparticles leads to the waveguiding effect along the beam
propagation direction inside KDP layers. The photoinduced
waveguiding effect reduces the scattering of light: at least by 30\%
for Pr-5 and by a factor of two for Pr-4 (see Table~2) as compared
with the nonresonant excitation case.

The Pr-4 sample possesses unique useful optical properties among the
rest investigated KDP crystals with incorporated TiO$_2$
nanoparticles: enhanced transmittance in the UV-visible range and
low scattering losses at the 532-nm resonant laser excitation.

\subsubsection*{3.1.3. Cone-shaped interference analysis}

Pure KDP belongs to uniaxial negative crystals. The incorporation of
TiO$_2$ nanoparticles produces additional stresses/strains in
crystals. Due to this fact, the crystals acquire the anomalous
biaxiality (AB) which can be measured for $Z$-cut plates within the
cone-shaped interference analysis. The technique is the following: a
sample is mounted between crossed polarizers, and it is illuminated
with a cone-shaped beam.

The cone-shaped interference images for $Z$-cut plates of pure
KDP(P) ({\it a}) and P-4 ({\it b}) samples are presented in Fig.~5.
For isotropic media like pure KDP(P), two isogyres overlap along the
optical axis and form the dark cross ({\it a}). The incorporation of
nanoparticles induces the AB effect and a lateral optical
anisotropy. The last one exhibits itself as the isogyre divergence
({\it b}). The extent of the AB effect is determined from the
magnitude of isogyre divergence (the angle 2$V$) in a cone-shaped
interference pattern.

The distance between the isogyres is independent of the rotational
orientation of a sample and depends on the 2$V$ value of a material.
When this 2$V$ value is large (about 40--50$^{\circ}$), two isogyres
are rarely  seen at a single image. However, for materials with a
weak AB effect, the 2$V$ value is small and can be measured directly
from the image.

\begin{table}[b]
\noindent\caption{Results of the cone-shaped interference analysis
for KDP(P) series}\vskip3mm\tabcolsep8pt

\noindent{\footnotesize\begin{tabular}{ccccccc}
  \hline
         \multicolumn{1}{c}{TiO$_2$ conc.,} & \multicolumn{4}{|c}{Intrinsic strain anomalies} & \multicolumn{1}{|c}{} \\

        \multicolumn{1}{c}{wt.\%}  & \multicolumn{4}{|c|}{(number of points)} & \multicolumn{1}{|c}{$2V_{\rm
        av.}$}
        \\ \cline{2-5}

          & \multicolumn{1}{|c}{0--$56'$} & \multicolumn{1}{|c}{5--$10'$} & \multicolumn{1}{|c}{10--$15'$} & \multicolumn{1}{|c}{15-$20'$} &\multicolumn{1}{|c}{} \\ \hline

        0 & 60 & 40 & -- & -- & 6 \\ 

        $10^{-5}$ & -- & -- & 80 & 20 & 12 \\ 

        $10^{-4}$ & -- & -- & 40 & 60 & 17 \\ \hline
 \end{tabular}}
\end{table}

Values of 2$V$ were measured for KDP(P) crystals over different
lateral points of samples. The results are summarized in Table~3.
They show that the strains in pure KDP single crystals are very
small. The manifestation of the AB effect enhances with increase in
the concentration of incorporated TiO$_2$ nanoparticles in the
matrix. However, even for $10^{-4}$ TiO$_2$ wt.\%, the value of 2$V$
is less than $20'$, which indicates a good optical quality of the
samples under study.

\begin{figure*}
\includegraphics[width=17cm]{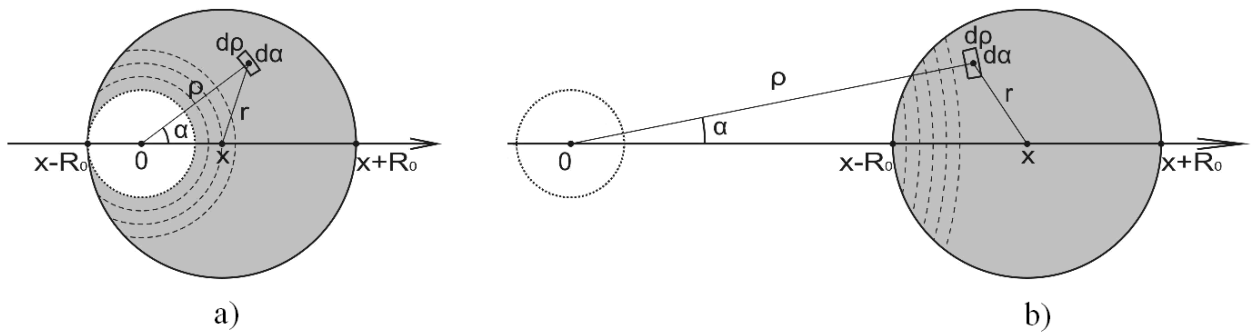}
\caption{Lines of equal radiation intensities on the lens: when the
laser beam axis is in ({\it a}) and beyond ({\it b}) the lens
aperture}
\end{figure*}

\subsection{The mathematical model}

As was shown above, the registered scattering indicatrices are
distorted by the presence of the lens, especially in areas 1 and 2
in Fig.~4. Let $\theta_0$ denote the current position of the lens
center. Then the registered dependences $P(\theta_0)$ are quite
precise at $\theta_0 > 2^{\circ}$. But, at small angles, they need
an additional processing. When the \textquotedblleft
wings\textquotedblright \space of indicatrices are compared on the
logarithmic scale, the central part $| \theta_0 | \leqslant
2^{\circ}$ should be given on a linear scale for convenience.

There are two possible ways to obtain the real view of the
indicatrix at $\theta_0 \leqslant 2^{\circ}$. The first way is
experimental (using an optical fiber spectrophotometer instead of a
CCD with a lens), and the second one is a way of mathematical
processing. The first way is rather complicated and requires a lot
of time to execute experiments. So it is reasonable to develop a
simulation procedure to solve this problem, and the current
mathematical model is the first step on this way. The verification
of the model was performed, by using a 532-nm continuous laser and
the series of KDP(P) crystals.

For isotropic samples, the intensity distribution of scattered
radiation is axially symmetric, so the lines with the same intensity
are circumferences of radii $\rho = L \sin {\theta}$ (Fig.~6), where
$\theta$ is the azimuth angle, and $L$ is the distance between the
sample and the lens (Fig.~1). As the registration unit moves along a
circle, it is reasonable to use a spherical coordinate system. But,
to process the data registered at the rotation angles $| \theta_0 |
\leqslant 2^{\circ}$ of a movable goniometer arm, a cylindric
coordinate system can be used. The coordinate origin $O$ is the
point where the laser beam axis hits the center of a lens at
$\theta_0 = 0$. Then, at $\theta_0 \neq 0$, the lens center
coordinate $x = L \sin {\theta_0} \approx L \theta_0$ (Fig.~6),
$\alpha$ is the polar angle, and $r$ is the distance from an
arbitrary point on the lens aperture to the lens center. The laser
beam axis in Fig.~6 is perpendicular to the plane of the figure.

For convenience, $u(x)$ in the mathematical part of the work means
the registered scattered power $\Delta P(\theta_0)$, and $\nu
(\rho)$ stands for the real scattered intensity $I_{\rm
scat}(\theta)$. Then the infinitesimal amount of a registered
radiation power at each point of the lens can be given by
\begin{equation}\label{E:diff}
 \Delta u = \nu (\rho) G(x,\rho,\alpha) \Delta S,
\end{equation}
where $G$ is the propagation function at a specific point on the
lens, and $\Delta S$ is an area element of the lens aperture
(Fig.~6).

The integration of Eq.~(2) over the lens aperture gives the
registered radiation power at a certain position $\theta_0$ of a CCD
with a lens. If we assume that the propagation function depends only
on the polar radius of a point on the lens $G = \hat{G}(r)$, we can
write
\begin{equation}\label{E:2int}
u(x) = \!\! \int \limits_{\rho_1 (x)}^{\rho_2 (x)} \!\! \rho \nu
(\rho) \!\!\! \int \limits_{\alpha_1 (x,\rho)}^{\alpha_2 (x,\rho)}
\!\!\! G(x,\rho,\alpha) d\rho d\alpha,~~~x \in [-a,a].
\end{equation}

\noindent Here, $G(x,\rho,\alpha) = \hat{G}(r)$ according to
\begin{equation}\label{E:cos}
r^2 = \rho^2 + x^2 - 2 \rho x \cos \alpha.
\end{equation}

According to the geometry of the experiment ($L$~= 200~mm, $a
\leqslant 7$~mm), Eqs.~(3)--(4) are written with an error less than
$0.08~\%$.

The dependence $\hat{G}(r)$ can be directly measured if an ideally
thin laser beam scans the lens along its diameter. In this case, at
a point $r = R_0$ ($R_0 = d/2$ is the lens radius), the abrupt drop
in a registered signal from a finite value to zero would be
observed. But, by mathematical reasons, $\hat{G} (r)$ should be
continuous at every point of $r \in [0, + \infty)$, and this abrupt
drop should be eliminated.

As the laser beam diameter $\delta \approx$ 0.9~mm is sufficiently
less than the lens size, the experimental data for a freely
propagating laser beam were taken as a propagation function. It can
be approximated with a continuous function, and its form is similar
to the real propagation function because $\delta \ll d$. The results
of the experimental measurements are shown in Fig.~7. As a
mathematical approximation, we use the function
\[ \hat{G} (r) = \chi (R_1^2 - r^2) (A \exp ((r/r_{A})^2) - \]
\begin{equation}
- (r/r_{B})^2 - C \exp (-\frac{(r/r_{C})^2}{1 - (r/r_{H})^2}) \chi
(r_{H}^2 - r^2)),
\end{equation}
where $R_1 = 5.509~{\rm mm},$
\[
\begin{array}{lll}
A = 2.223, & r_{A} = 5.808~{\rm mm}, & C = 1.277, \\
r_{B} = 2.357~{\rm mm}, & r_{C} = 3.574~{\rm mm}, & r_{H} = 4.711~{\rm mm}, \\
\end{array}
\]
and $\chi (s)$ is the Heaviside function equal to unity for $s
\geqslant 0$ and equal to zero for $s < 0$.

In Eq.~(5), we have $\hat{G} (r) \equiv 0$ for $r > R_1$. Here, $R_1
> R_0$, because the beam radius is non-zero. The relative
approximation error is 3.6\%, while the relative experimental error
is $\approx 5\%$.

\begin{figure}
\includegraphics[width=\column]{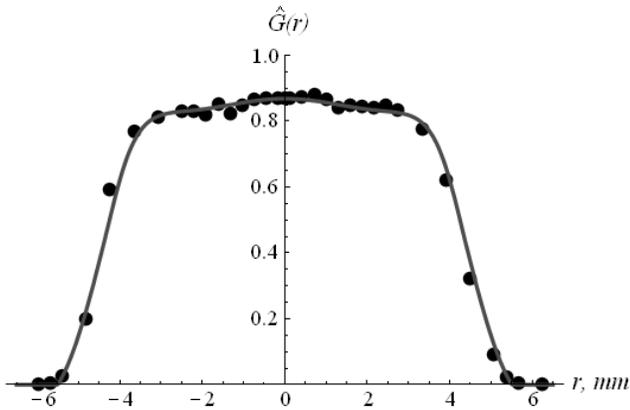}
\caption{Propagation function of the experimental setup (solid line)
and experimental data for a freely propagated beam (dots)}
\end{figure}

From the symmetry in Fig.~6, it is clear that, in Eq.~(3), $\alpha_1
(x,\rho) = - \alpha_2 (x,\rho)$. Using Eq.~(4), we can write
\begin{equation}
\alpha_2 (x,\rho) = \left\{
\begin{array}{ll}
\arccos \left( \frac{\rho^2 + x^2 - R_1^2}{2 \rho x} \right), & \rho + x > R_1, \\
\pi, & \rho + x \leqslant R_1.
\end{array}
\right.
\end{equation}

\noindent Using Eq.~(6) and the function $G(x,\rho,\alpha)$, one can
found the inner integral in Eq.~(3). As $G(x,\rho,\alpha)$ is a
symmetric function, it can be written as
\begin{equation}
K(x,\rho) = \frac{1}{2} \!\! \int \limits_{\alpha_1
(x,\rho)}^{\alpha_2 (x,\rho)} \!\!\! G(x,\rho,\alpha) d\alpha =
\!\!\! \int \limits_{0}^{\alpha_2 (x,\rho)} \!\!\!\!
G(x,\rho,\alpha) d\alpha.
\end{equation}

\noindent Equation (7) enables us to write the integral relationship
Eq.~(3) in the form
\begin{equation}
u(x) = 2 \!\! \int \limits_{\rho_1 (x)}^{\rho_2 (x)} \!\!\! \rho \nu
(\rho) K(x,\rho) d\rho,~~~~~x \in [-a,a].
\end{equation}

As for determining the integration limits $\rho_1 (x)$ and $\rho_2
(x)$, two positions of the lens with respect to the center of a
laser beam are possible. In the first case, the center of the beam
lies inside the circle of the radius $R_1$ (Fig.~6,{\it a}, $x
\leqslant R_1$). In the second case, the center of the beam is
outside the circle (Fig.~6,{\it b}, $x > R_1$). At first, we discuss
both of the cases apart.

1) $x \leqslant R_1$.

According to Eq.~(6), in this case, the lines of equal radiation
intensities inside the circle with $R_1$ may be of two types:

\noindent a) Circumferences of radii $0 \leqslant \rho \leqslant
(R_1 - x)$. Then $\alpha_2 (x,\rho) = \pi$ and
\[
K_0 (x,\rho) = \! \int \limits_{0}^{\pi} \! G(x,\rho,\alpha)
d\alpha,
\]
\begin{equation}
u(x) = 2 \!\! \int \limits_{0}^{R_1 - x} \!\!\! \rho \nu (\rho) K_0
(x,\rho) d\rho.
\end{equation}

\noindent b) Arcs of the circumferences $(R_1 - x) < \rho < (R_1 +
x)$. Then $\alpha_2 (x,\rho) < \pi$ according to Eq.~(6) and
\[
K_1 (x,\rho) =  \int \limits_{0}^{\alpha_2 (x,\rho)}
G(x,\rho,\alpha) d\alpha, \]
\begin{equation}
u(x) = 2 \!\! \int \limits_{R_1 - x}^{R_1 + x} \!\!\! \rho \nu
(\rho) K_1 (x,\rho) d\rho.
\end{equation}

The functions $K_0 (x,\rho)$ and $K_1 (x,\rho)$ are presented in
Fig.~8. They can be united into a 2D continuous function $K
(x,\rho)$.

2) $x > R_1$.

In this case, the lines of equal radiation intensities are also arcs
of the circumferences with radii $(x - R_1) < \rho < (x + R_1)$.
Here, $u(x)$ can be given identically to Eq.~(10), but with
different integration limits. The total integration relation Eq.~(8)
can be written using Eqs.~(9)--(10) and Eq.~(5) as follows:
\begin{equation}
u(x) = 2 \!\! \int \limits_{0}^{2 R_1} \!\! \rho \nu (\rho)
K(x,\rho) d\rho,~~~~~x \in [-a,a],
\end{equation}
where $K(x,\rho)$ is the union of the functions $K_0 (x,\rho)$ and
$K_1 (x,\rho)$ (Fig.~8).

There are two ways to interpret Eq.~(11). When the real intensity
distribution of the scattered light $\nu (\rho)$ is known, Eq.~(11)
becomes the integration formula one can use to mathematically obtain
the resulting data of the experiment with certain parameters of the
laser setup. In this case, the direct problem is solved, and no
computational difficulties arise.

Another situation appears at solving the inverse problem. Here,
$u(x)$ is known, and the task is to reconstruct the real intensity
distribution $\nu (\rho)$ of the radiation scattered by the sample.
Then Eq.~(11) serves as the Fredholm first-kind integral equation,
and its solution is a solution of an incorrect problem. In this
case, special regularization methods are required
[6,~7].\looseness=1

\begin{figure}
\includegraphics[width=\column]{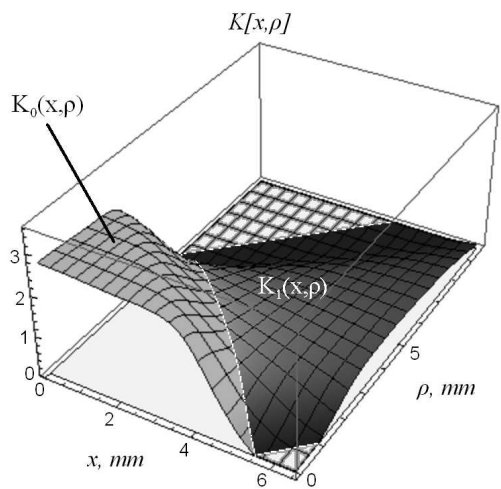}
\caption{Form of the integrand function $K(x,\rho)$}
\end{figure}

At the stage of verification of the consistency of the mathematical
model of the experimental laser setup for the optical diagnostics of
low-dispersion samples, it is suitable to have the experimental data
for $\nu (\rho)$ and $u(x)$. For this purpose, we studied

a) Freely propagated 532-nm laser beam;

b) Laser beam passes through a pure KDP(P) crystal;

c) Laser beam passes through a KDP crystal doped with TiO$_2$
nanoparticles (sample P-4).

To experimentally obtain the data on the real intensity distribution
$\nu (\rho)$, we used the experimental setup with an optical-fiber
(diameter $d = 410~\mu$m) spectrophotometer instead of a CCD with a
lens. Results of the experiment reveal that, in all three
investigated cases, the propagating laser beam has the Gaussian
intensity distribution
\begin{equation}
\nu (\rho) = \exp (- (\rho / \rho_0)^2).
\end{equation}

A beam size reduction in the registration plane after the
transmission of samples at 532 nm was shown ($\rho_0$ = 0.447~mm in
the case of free propagation of the beam). In case of KDP with
TiO$_2$ (0.421~mm), it is less than that for pure KDP (0.432~mm).
This can be explained by the photoinduced self-focusing effect in
these samples at the 532-nm CW laser excitation, which will be
published elsewhere in detail. This result proves that knowing the
real scattering indicatrix can be essential for the optical
characterization of a sample.

In Fig.~9,{\it a}, the experimental data (dots) and the calculated
$u(x)$ dependence (solid line) for P-4 sample are presented. We also
show the precise scattering data for P-4 sample (dashed line,
normalized to $\sim 5 \times 10^{-3}$ for convenience).
Figure~9,{\it b} shows the difference between the calculated
indicatrices for a freely propagated laser beam, pure KDP(P), and
P-4 sample. The result presented is in good agreement with spectral
data (Table~1). The dependences $u(x)$ for all the samples are
obtained, by using Eq.~(11) and functions (12), the approximation
error being less than $5\%$.

Based on the results of the work, it can be stated that the
mathematical model of the experimental laser setup for the optical
diagnostics of low-dispersion samples is consistent, and it can
serve as a base for developing a way to solve the inverse problem,
i.e. to get the real scattering data $I_{\rm scat}(\theta)$ out of
the registered $\Delta P(\theta_0)$ using mathematical calculations.

\begin{figure}
\includegraphics[width=\column]{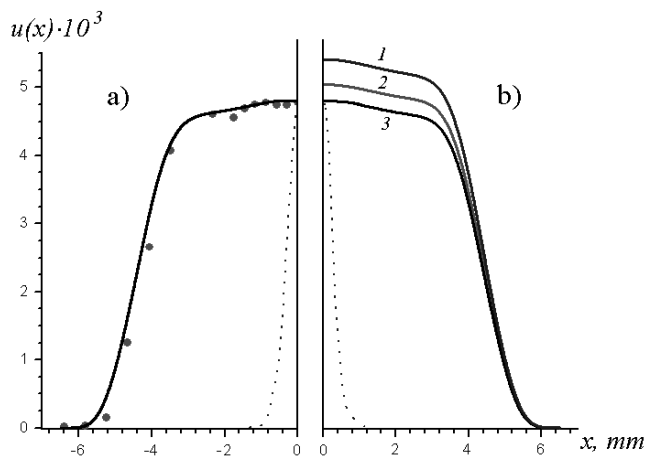}
\caption{Comparison of the experimental data at 532 nm and the
simulation results. ({\it a}) Dots -- experimental data for P-4
sample, solid line -- simulation, dashed line --- precise
measurements (normalized for convenience). ({\it b}) Simulation
results: {\it 1} -- freely propagated laser beam, {\it 2} -- KDP(P)
pure, {\it 3} -- P-4 sample}
\end{figure}

\section{Summary}

In this work, the optical characterization of series of KDP crystals
doped with TiO$_2$ nanoparticles (anatase modification) in various
concentrations was performed. The characterization includes
measurements of scattering indicatrices and optical transmission
spectra and the cone-shaped interference method.

The investigations show that the incorporation of TiO$_2$
nanoparticles slightly changes the optical quality of KDP crystals.
All of the characterization results contribute to a good optical
quality of the investigated samples: transmittances in the visible
range $\sim~90\%$, scattering losses $< 3\%$, and the anomalous
biaxiality  values $2V < 20'$.

A mathematical model of the experimental setup was developed in
order to enhance the angular resolution in the scattering
indicatrices at angles $\theta \leqslant 2^{\circ}$, where the
registered data are blurred due to the presence of a lens in the
measuring circuit. The model was verified in the direct way and
showed itself to be consistent with an error less than $5\%$.

\vskip3mm The work was supported by STCU-NASU grant No.~4956.

\end{document}